\documentclass[twocolumn,superscriptaddress]{revtex4}

\usepackage{bbold}
\usepackage{mathptmx}
\usepackage{subfigure}
\usepackage{psfrag,graphicx}
\usepackage{dcolumn}
\usepackage{amsmath,amssymb}
\usepackage{bm}
\usepackage{color}
\usepackage{latexsym}
\usepackage{epstopdf}
\usepackage{color}
\usepackage[english]{babel}
\usepackage{latexsym}
\usepackage{psfrag,graphicx}
\usepackage{subfigure}
\usepackage{amsmath}
\usepackage{amssymb}
\usepackage{amsfonts}
\usepackage{bm}
\usepackage{natbib}
\usepackage{epstopdf}
\usepackage{appendix}
\usepackage{rotating}

\definecolor{mygrey}{gray}{0.35}
\definecolor{myblue}{rgb}{0.2,0.2,0.8}
\definecolor{myzard}{cmyk}{0,0,0.05,0}
\definecolor{mywhite}{rgb}{1,1,1}
\definecolor{mywhite}{rgb}{1,1,1}
\definecolor{myred}{rgb}{1,0.,0.3}

\usepackage[colorlinks=true,citecolor=myblue,linkcolor=myred]{hyperref}

\def\ba{\begin{align}}
\def\enda{\end{align}}
\def\bi{\begin{itemize}}
\def\ei{\end{itemize}}

\def\be{\begin{equation}}
\def\ee{\end{equation}}
\def\bea{\begin{eqnarray}}
\def\eea{\end{eqnarray}}
\def\bse{\begin{subequations}}
\def\ese{\end{subequations}}

\newcommand{\ket}[1]{|{#1}\rangle}                       
\newcommand{\bra}[1]{\langle {#1}|}                      
\newcommand{\average}[1]{\langle {#1} \rangle}           

\newcommand{\Ignore}[1]{ }

\begin{document}

\preprint{APS/123-QED}

\title{Time evolution of a pair of distinguishable interacting spins subjected to controllable and noisy magnetic fields}

\author{R. Grimaudo}
\address{ Dipartimento di Fisica e Chimica dell'Universit\`a di Palermo, Via Archirafi, 36, I-90123 Palermo, Italy.}
\address{I.N.F.N., Sezione di Catania, Catania, Italy.}

\author{Yu. Belousov}
\affiliation{Moscow Institute of Physics and Technology, 141700, Dolgoprudny, Institutsky lane 9, Russia.}

\author{H. Nakazato}
\affiliation{
 Department of Physics, Waseda University, Tokyo 169-8555, Japan.\\
}

\author{A. Messina}%
\address{I.N.F.N., Sezione di Catania, Catania, Italy.}
\address{ Dipartimento di Matematica ed Informatica dell'Universit\`a di Palermo, Via Archirafi, 34, I-90123 Palermo, Italy.}%




\date{\today}

\begin{abstract}
The quantum dynamics of a $\hat{\mathbf{J}}^2=(\hat{\mathbf{j}}_1+\hat{\mathbf{j}}_2)^2$-conserving Hamiltonian model describing two coupled spins $\hat{\mathbf{j}}_1$ and $\hat{\mathbf{j}}_2$ under controllable and fluctuating time-dependent magnetic fields is investigated.
Each eigenspace of $\hat{\mathbf{J}}^2$ is dynamically invariant and the Hamiltonian of the total system restricted to any one of such $(j_1+j_2)-|j_1-j_2|+1$ eigenspaces, possesses the SU(2) structure of the Hamiltonian of a single fictitious spin acted upon by the total magnetic field. 
We show that such a reducibility holds regardless of the time dependence of the externally applied field as well as of the statistical properties of the noise, here represented as a classical fluctuating magnetic field.
The time evolution of the joint transition probabilities of the two spins $\hat{\mathbf{j}}_1$ and $\hat{\mathbf{j}}_2$ between two prefixed factorized states is examined, bringing to light peculiar dynamical properties of the system under scrutiny.
When the noise-induced non-unitary dynamics of the two coupled spins is properly taken into account, analytical expressions for the joint Landau-Zener transition probabilities are reported.
The possibility of extending the applicability of our results to other time-dependent spin models is pointed out. 
\end{abstract}

\maketitle


\section{Introduction}

Almost eighty years ago I. I. Rabi published two seminal papers \cite{Rabi 1937,Rabi 1954} where he solved the quantum dynamics of a semi-classical system consisting of a spin 1/2 immersed in the following time-dependent magnetic field
\begin{equation}\label{RMF}
\textbf{B} = B_\perp(\cos(\omega t) \textbf{c}_x+\sin(\omega t) \textbf{c}_y)+B_0 \textbf{c}_z,
\end{equation}
constant in magnitude and precessing around the $z$-axis with angular frequency $\omega$.
Here $\textbf{c}_x$, $\textbf{c}_y$ and $\textbf{c}_z$ are fixed unit vectors in the laboratory frame.
The exact knowledge of the unitary time evolution in such a case provides the basic building block for exactly solving the quantum dynamics of a spin $\hat{\mathbf{J}}$, of arbitrary magnitude $J$, in the same magnetic field adopted by Rabi \cite{Hioe}.

The group-theoretical based protocol enabling the construction of the evolution operator of the spin $\hat{\mathbf{J}}$ from that relative to the spin 1/2, holds its validity whatever the time-dependent magnetic field is \cite{Hioe}.
Of course, the usefulness of such a protocol depends on our ability to exactly solve the SU(2) time-dependent problem of a spin 1/2 \cite{Hioe}.
Thus, it is of relevance that, quite recently, a new strategy to single out controllable time dependent magnetic fields for which exact analytical solutions of the relative Schroedinger-Liouville equations for the SU(2) evolution operator $U$, has been reported \cite{Bagrov, KunaNaudts, Das Sarma, Mess-Nak, GMN, MGMN}.

In this paper we call Generalized Rabi System (GRS) an SU(2) system consisting of a spin $\hat{\mathbf{J}}$ in a controlled time-dependent magnetic field.
The scope of this paper is to investigate the mutual dynamical influence between two distinguishable GRSs coupled by an isotropic Heisenberg exchange term.
Environmental noise is even incorporated in our theory adding the action of a classical fluctuating magnetic field on the two-spin system.
Notice that there exist several physical scenarios of experimental interest wherein the coupling between two GRSs cannot be ignored.

For example, an isolated dimeric unit of ions, each one exhibiting an effective spin $\hat{\mathbf{J}}$, may be regarded, as a system of two interacting spins.
For some compounds of dimeric units it has been experimentally proven that neglecting the couplings between spins in neighbouring units is legitimate, implying that the quantum dynamics of the same compound may be derived from that of a single dimer \cite{Calvo1,Calvo2}.
Experimental and theoretical investigations on biradical compounds provide a further example of a physical system describable in terms of two interacting spins.
In a liquid solution a compound of biradicals can be described by a symmetric spin Hamiltonian model \cite{Bolton}.
Research activity involving biradical compound systems run from  high controllable low dimensional quantum magnets realization to the study of Bose-Einstein condensation phenomena for magnetic excitations \cite{Borozdina}.
In the area of quantum computing, finally, spin Hamiltonian models describing the quantum dynamics of two electron spins in a double quantum dot \cite{Hayashy, Hu, Gorman, Petta, Mason} or in a double quantum well \cite{Anderlini1,Anderlini2}, furnish a theoretical basis for manipulating two-electron based qubits.

To achieve realistic descriptions of these spin systems, environmental disturbs cannot be neglected \cite{Das Sarma Nature, Das Sarma PRB, Das Sarma PRA, Ryzhov, Maune, Foletti1,Foletti2,Foletti3}.
Consider for example a couple of interacting spins hosted in a real or artificial lattice.
The magnetic field acting upon such a spin system results from a controllable, generally time-dependent, contribution and a random one originated by interactions with the environmental nuclear spin bath around the system under scrutiny.
In this paper we describe for simplicity this random fluctuating magnetic field as a classical random field characterized by statistical properties mimicking the quantum fluctuations of the previously mentioned bath \cite{Pokrovsky1,Pokrovsky2,Kenmoe1,Kenmoe2}.
Postulating that this field acting upon the spins stems from mechanisms independent of the applied controllable field, in this paper we sketch a very basic experimental scheme for checking the classical versus quantum description of the random field.
Our approach to the treatment of noise effects is illustrated by a specific scenario wherein two spin 1/2's are subjected to a Landau-Zener time-dependent controllable magnetic field.

The main result of this paper is to show the symmetry-based reducibility of the quantum dynamics of our two-spin system to that of a single spin in presence of both time-dependent controllable and random fluctuating magnetic fields.

The paper is organized as follows.
The two-spin time-dependent Hamiltonian model and general features of the generated evolution operator are reported in Sec. \eqref{TDM and TEO}, while applications leading to some peculiar quantum dynamical results are discussed in Sec. \eqref{QM and IFE}.
Three exemplary cases are developed in Sec. \eqref{Ex Cases} to illustrate our approach while the time behaviour of magnetization and of other physical quantities of experimental interest are presented in the Sec. \eqref{Int Evid}.
More realistic physical scenarios for our two-spin system are taken into consideration in Sec. \eqref{Noise}, by introducing into the Hamiltonian a classical fast Gaussian noisy term.
The joint Landau-Zener transition probabilities are explicitly given in the case of two spin 1/2's.
Some conclusive remarks are finally pointed out in the last section.

\section{The time-dependent Hamiltonian model and the evolution operator structure}\label{TDM and TEO}

Our physical system consists of two independent, localized and distinguishable spins of different value and physical nature, in general, represented by their relative spin operators  $\hat{\textbf{j}}_1$ and $\hat{\textbf{j}}_2$, respectively, with $\hat{\textbf{j}}_i \equiv (j_i^x,j_i^y,j_i^z)$ ($i=1,2$).
By definition $[\hat{j}_1^\alpha,\hat{j}_2^\beta]=0$ ($\alpha,\beta=x,y,z$) and $\hat{\textbf{j}}_i \wedge \hat{\textbf{j}}_i = i \hbar \hat{\textbf{j}}_i$.
The $i$-th spin is subjected to the local external controllable time-dependent magnetic field
\begin{equation}
\textbf{B}_i(t) = B_i^x(t) \textbf{c}_x+B_i^y(t) \textbf{c}_y+B_i^z(t) \textbf{c}_z,
\end{equation}
such that
\begin{equation}\label{Def Omega}
-\gamma_1\textbf{B}_1(t)=-\gamma_2\textbf{B}_2(t)\equiv \bm{\Omega}(t),
\end{equation}
$\gamma_i=g_i\mu_0$ being the magnetic moment associated to the $i$-th spin, with $g_i$ the appropriate Land\'e factor, and $\mu_0$ the appropriate Bohr magneton.
We observe that $\textbf{B}_1(t)$ and $\textbf{B}_2(t)$ are parallel (anti-parallel) if $\gamma_1\gamma_2>0$ ($<0$).

Condition \eqref{Def Omega} means that the two spins exhibit the same Zeeman spitting.
The possibility of such a control of the magnetic field acting individually on each spin is in the grasp of experimentalists as realized in a double quantum dot system \cite{Das Sarma Nature, Das Sarma PRB}. 

Let us suppose that the two spins are in addition coupled via a ferromagnetic or anti-ferromagnetic isotropic Heisenberg interaction of strength $\lambda$, so that the corresponding Hamiltonian model may be written down as follows (from now on we set $\hbar=1$):
\begin{equation}
H(t)=H_0(t)+H_I
\end{equation}
with
\begin{equation}\label{Hamiltonian}
H_0(t)=\sum_{i=1}^2 \bm{\Omega}(t) \cdot \hat{\textbf{j}}_i, \quad H_I= -\lambda \hat{\textbf{j}}_1 \cdot \hat{\textbf{j}}_2,
\end{equation}
acting upon the $(2j_1+1)(2j_2+1)$-dimensional Hilbert space $\mathcal{H}$ of the two spins.

We emphasize that recent experimental advances in the area of $^{28}$Si-based solid state quantum computing makes our Hamiltonian model \eqref{Hamiltonian} of some help to represent such a physical scenario \cite{Das Sarma PRB}.
In addition biradical compounds in liquid phase provide another interesting experimental situation usefully describable making use of our noiseless model \cite{Bolton}.

To proceed in the analysis of the model it is useful to rewrite the Hamiltonian in terms of the total spin angular momentum operator $\hat{\textbf{J}}=\hat{\textbf{j}}_1+\hat{\textbf{j}}_2$, getting
\begin{equation}\label{H in terms of J^2}
H(t)=\bm{\Omega}(t) \cdot \hat{\textbf{J}} - {\lambda \over 2} \hat{\textbf{J}}^2 + K
\end{equation}
where $K \equiv {\lambda \over 2}(\hat{\textbf{j}}_1^2+\hat{\textbf{j}}_2^2)$ is proportional to the identity operator. 
Equation \eqref{H in terms of J^2} clearly shows that our time-dependent Hamiltonian $H(t)$ commutes with $\hat{\mathbf{J}}^2 = (\hat{\mathbf{j}}_1+\hat{\mathbf{j}}_2)^2 $ and this implies that $\text{Tr}\{ \rho(t) \mathbf{\hat{J}}^2 \} = \text{Tr}\{ \rho(0) \mathbf{\hat{J}}^2 \}$ at any time instant.
Here $\rho(t) = U(t)\rho(0)U^\dagger(t)$, $U(t)$ being the unitary time evolution operator fulfilling the Cauchy problem
\begin{equation}\label{Cauchy problem U}
i\dot{U}(t)=H(t) U(t) \qquad U(0)=\mathbb{1}
\end{equation}
and $\rho(0)$ the initial density matrix of the two spins.

The conservation of $\mathbf{\hat{J}}^2$ leads to the existence of $(j_1+j_2)-|j_1-j_2|+1$ 
orthogonal, dynamically invariant subspaces $\mathcal{H}^{(j)}$ such that
\begin{equation}
\mathcal{H}=\bigoplus_{j=|j_1-j_2|}^{j_1+j_2}\mathcal{H}^{(j)}
\end{equation}
$\mathcal{H}^{(j)}$ denoting the invariant $(2j+1)$-dimensional subspaces of $\mathbf{\hat{J}}^2$ pertaining to its eigenvalue $j(j+1)$.
The Hamiltonian operator may be written as
\begin{equation}\label{H direct sum}
H(t)=\bigoplus_{j=|j_1-j_2|}^{j_1+j_2} H^{(j)}(t), 
\end{equation}
and accordingly generates the time evolution operator, solution of the Cauchy problem defined in Eq. \eqref{Cauchy problem U} in the form
\begin{equation}\label{U direct sum}
U(t)=\bigoplus_{j=|j_1-j_2|}^{j_1+j_2} U^{(j)}(t). 
\end{equation}
$H^{(j)}(t)$ is the effective Hamiltonian of the two spins governing their dynamics in the $(2j+1)$-dimensional dynamically invariant subspace $\mathcal{H}^{(j)}$ of $H(t)$, whereas $U^{(j)}(t)$ is the related time evolution operator, solution of the (restricted) Cauchy problem 
\begin{equation}\label{Cauchy problem Uj}
i\dot{U}^{(j)}(t)=H^{(j)}(t)U^{(j)}(t), \qquad U^{(j)}(0)=\mathbb{1}^{(j)},
\end{equation}
$\mathbb{1}^{(j)}$ being the identity operator in $\mathcal{H}^{(j)}$.

Since the term $K'\equiv -{\lambda \over 2} \hat{\textbf{J}}^2+K$ is proportional to $\mathbb{1}^{(j)}$ in $\mathcal{H}^{(j)}$, whatever $j$ is, the effective Hamiltonian $H^{(j)}(t)$ governing the dynamics in $\mathcal{H}^{(j)}$ may be written as
\begin{equation}\label{Hj}
H^{(j)}(t)=\bm{\Omega}(t) \cdot \hat{\textbf{j}} + K',
\end{equation} 
which, formally, is the Hamiltonian of a fictitious spin  $\hat{\textbf{j}}$, with spin angular momentum $j$ and magnetic moment $\gamma_1$, subjected to the time-dependent magnetic field $\mathbf{B}_1(t)$.
Of course, due to Eq. \eqref{Def Omega} and Eq. \eqref{Hj}, in this scenario $\gamma_1$ and $\mathbf{B}_1(t)$ may be replaced by $\gamma_2$ and $\mathbf{B}_2(t)$, respectively.
This means that each effective Hamiltonian $H^{(j)}(t)$ possesses an SU(2)-symmetry structure and the related time evolution operator $U^{(j)}(t)$ may be expressed \cite{Hamermesh, Weissbluth} in terms of the two time-dependent complex-valued functions, $a=a(t)$ and $b=b(t)$, entries of the evolution operator
\begin{equation}\label{TEO spin 1/2}
U^{(1/2)} (t)=
\begin{pmatrix}
a & b \\
-b^* & a^*
\end{pmatrix},
\quad
|a|^2+|b|^2=1,
\end{equation}
i.e. the solution of the Liouville-Cauchy problem \eqref{Cauchy problem Uj} with $j=1/2$ and $H^{(1/2)}=\bm{\Omega}(t) \cdot {\hat{\bm \sigma} \over 2}$.
The Pauli vector is defined as $\hat{\bm \sigma} = \sigma^x \mathbf{c}_1+ \sigma^y \mathbf{c}_2+ \sigma^z \mathbf{c}_3$, $\sigma^x$, $\sigma^y$ and $\sigma^z$ being the Pauli matrices.
The entries of the matrix $U^{(j)}(t)$ in the standard ordered basis of the eigenstates of the third component of the fictitious spin $j$: $\{ \ket{m}, m=j,j-1, \dots, -j \}$, may be cast as follows \cite{Weissbluth} (time dependence is suppressed)
\begin{equation}\label{Ujmm'}
U^{(j)}_{m,m'}(a,b)=e^{-i{K'} t}\sum_\mu C^{(j)}_{m,m'} a^{j+m'-\mu}(a^*)^{j-m-\mu}b^{m-m'+\mu}(b^*)^\mu,
\end{equation}
where \cite{Weissbluth}
\begin{equation}\label{Cjmm'}
C^{(j)}_{m,m'} = (-1)^\mu {\sqrt{(j+m)!(j-m)!(j+m')!(j-m')!} \over \mu!(j+m'-\mu)!(j-m-\mu)!(m-m'+\mu)!}.
\end{equation}
We point out that, whatever $m$ and $m'$ are, the summation, formally a series generated by $\mu$ running over the integer set $\mathbb{Z}$, is a finite sum, generated by all the values of $\mu$ satisfying the condition $\text{Max}[0,m'-m]\leq\mu\leq\text{Min}[j+m',j-m]$.
It is possible to convince oneself that, defined in this manner, $\bigl|U^{(j)}_{m,m'}(a,b)\bigr|^2$ represents the probability to find the $N$-level system in the state with $z$-projection $m$ when it is initially prepared in the state with $z$-projection $m'$.
Summing up, Eqs. \eqref{Ujmm'} and \eqref{Cjmm'} provide the solution of the Cauchy problem defined by Eq. \eqref{Cauchy problem Uj} which, in turn and in view of Eq. \eqref{U direct sum}, enables us to write down the exact time evolution operator solution of our main problem as defined by Eq. \eqref{Cauchy problem U}.

We emphasize that the possibility of expressing $U(t)$ in terms of only two time-dependent complex-valued functions may be traced back to the existence of SU(2) structures nested in the Hamiltonian model given by Eq. \eqref{Hamiltonian}.
Such a property is a direct consequence of the symmetries possessed by the Hamiltonian model and paves the way to the exact determination of the evolution operator $U(t)$ generated by $H$.

 
This approach may be successfully exploited when $\bm{\Omega}(t)$ is such to allow the construction of explicit expression for $a(t)$ and $b(t)$ in a given specific physical situation.
For example, when $\bm{\Omega}(t)$ coincides with that considered originally by Rabi \cite{Rabi 1937}, we are in condition to construct the explicit form of the evolution operator \cite{Rabi 1937, Rabi 1945, Rabi 1954} generated by the correspondent $H$ given in Eq. \eqref{Hamiltonian} and as a consequence to investigate any aspect of the related quantum dynamics.
It is thus of relevance that recently other SU(2) time-dependent scenarios have been proposed and exactly solved \cite{Bagrov, KunaNaudts, Mess-Nak, Das Sarma, GMN, MGMN}, with application to interacting spin systems \cite{GMN,GMIV}.
This circumstance opens the possibility of applying the approach reported in this paper to several possible other scenarios of experimental interest, different from the one originally considered by Rabi.
The exact knowledge of how two-spin systems evolve under controllable time-dependent magnetic fields might be exploited to comply, on demand, with technological needs or experimental requests.

\section{Quantum dynamics of two coupled GRSs}\label{QM and IFE}

The main scope of this paper is to investigate possible effects of the exchange interaction $H_I$ between $\hat{\mathbf{j}}_1$ and $\hat{\mathbf{j}}_2$ on the quantum dynamics each spin would experience if $\lambda$ were absent.
Since the problem of a spin $\hat{\mathbf{J}}$ in a time-dependent magnetic field may be considered as a direct generalization of the problem of that treated by Rabi in his papers published in 1937 \cite{Rabi 1937} and 1954 \cite{Rabi 1954}, we are going to study the reciprocal influence of two generalized semiclassical Rabi systems stemming from their exchange coupling.

Let us denote by $\ket{m_i}$, $m_i=j_i,j_i-1,\dots,-j_i$, a generic eigenstate of $\hat{j}_i^z$ ($i=1,2$).
Suppose our two-spin system prepared in the state $\ket{j_1,j_2} \equiv \ket{j_1}\ket{j_2}$ belonging to $\mathcal{H}^{(j_1+j_2)}$.
The probability $P_{j_1,j_2}^{-j_1,-j_2}(t)$ of finding the compound system in the state $\ket{-j_1,-j_2} \in \mathcal{H}^{(j_1+j_2)}$ at any time instant may be expressed as 
\begin{equation}\label{P jj -j-j}
\begin{aligned}
P_{j_1,j_2}^{-j_1,-j_2}(t) &= \bigl|U_{-j,j}^{(j)}(a,b)\bigr|^2 = \\
 & = \bigl|[-b^*]^{2j} \bigr|^2 = |b|^{4(j_1+j_2)},
\end{aligned}
\end{equation}
in view of Eq. \eqref{Ujmm'}, with $j=j_1+j_2$.

This result means that, preparing the two spins in the factorized state $\ket{j_1,j_2}$, the probability of finding the system in the factorized state $\ket{-j_1,-j_2}$ is less than or equal to $|b|^4$ that is the probability up-down we would get when two non-interacting spin 1/2's experience the same $\bm{\Omega}(t)$ acting upon the two spins $\hat{\mathbf{j}}_1$ and $\hat{\mathbf{j}}_2$.
We emphasize that this effect is related to the increased dimension, from 2 to $2j+1$, of the Hilbert subspace where the time evolution takes place and in addition that such behaviour does not depend on the specific time dependence of $\bm{\Omega}(t)$.
Figure 1 illustrates this result for different values of $j_1$ and $j_2$, assuming $\bm{\Omega}(t)$ in accordance with Eq. \eqref{RMF}.
This is the semiclassical Rabi problem in the resonance condition whose exact solution may be analytically derived \cite{Rabi 1937, Rabi 1945, Rabi 1954}.
The plots are reported against the dimensionless time $\tau=\lambda_R t$, with $\lambda_R={\gamma B_\perp \over 2}$. 
\begin{figure}[tbph]
\centering
{\includegraphics[width=\columnwidth]{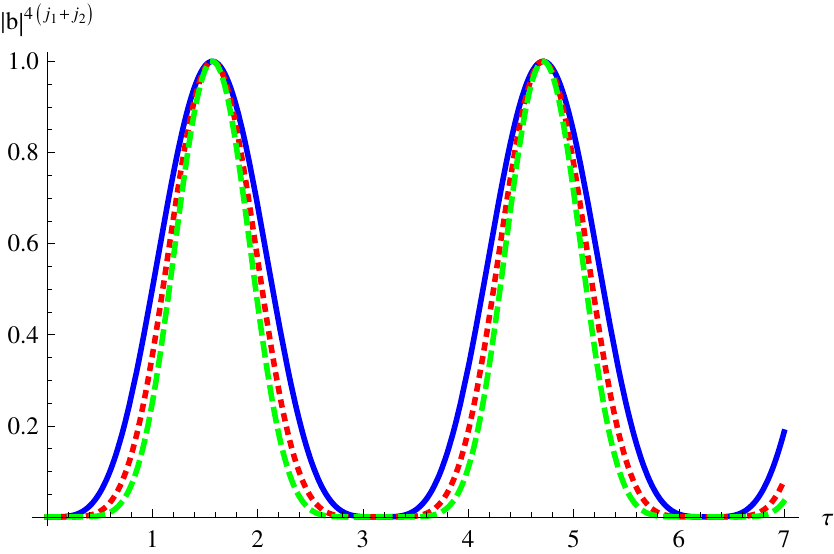} }
\caption{Plot of Eq. \eqref{P jj -j-j} with $j_1=j_2=1/2$ (blue full), $j_1=2j_2=1$ (red dotted), $j_1=j_2=1$ (green dashed), for the Rabi scenario \cite{Rabi 1937} characterized by the time-dependent magnetic field \eqref{RMF}; $\tau$ is the dimensionless time defined as $\tau=\lambda_R t$ with $\lambda_R=\gamma B_\perp$ in the resonance condition.}
\end{figure}

It is possible to find a physical reason at the basis of the previous result by bringing to light remarkable features characterizing the quantum dynamics of our two-spin systems by considering the reduced dynamics of the two subsystems of interest.

\Ignore{
First of all we want to point out the following aspect.
The squared modulus of the matrix elements $U^{(j)}_{m,m'}(a,b)$ of the time evolution operator $U^{(j)}$ give us, thus, the (fictitious) probability to find the fictitious spin angular momentum $\hat{\mathbf{j}}$ in the state characterized by a projection on the $z$-axis equal to $m'$, having been initially prepared in the state with $z$ projection equal to $m$.
It is important to underline, then, that if we want to read physically this probability we have to interpret it in terms of the states of the real system of the two interacting spins.
This can be done by basing on the underlying mapping between the states of the fictitious spin $j$ and those of the two spins spanning the related $\hat{\mathbf{J}}^2$ dynamically invariant subspace, the latter obtainable from the standard basis states through the Clebsh-Gordan matrix transformation.}
The Liouville-Cauchy problem governing the time evolution of a generic state $\rho(0)$ of the two-spin system is
\begin{equation} \label{L-C problem}
i\dot{\rho} = [H,\rho].
\end{equation}
In view of Eq. \eqref{U direct sum}, its solution $\rho(t)=U(t) \rho(0) U^\dagger(t)$ is determined after solving the following Cauchy problem for a spin 1/2
\begin{equation}\label{Spin 1/2 LE}
i\dot{U}^{(1/2)}=H^{(1/2)}U^{(1/2)}, \quad U^{(1/2)}(0)=\mathbb{1}^{(1/2)},
\end{equation}
where $H^{(1/2)}=\bm{\Omega}(t) \cdot \hat{\textbf{s}}$, with $\hat{\textbf{s}}={1\over 2} \hat{\bm{\sigma}}$.

The time evolution of the reduced density matrix of the $i$-th spin is related to the solution $\rho(t)$ of the Liouville-Cauchy problem \eqref{L-C problem} as follows
\begin{equation}\label{Reduce Eq. of Motion}
i\dot{\rho}_i = [H_i,\rho_i] + \text{Tr}_{k \neq i}\{[H_{I},\rho]\},
\end{equation}
where $H_i=\bm{\Omega}(t) \cdot \hat{\textbf{j}}_i$, $H_{I}$ is defined in Eq. \eqref{Hamiltonian} and $\rho_i(t)$ satisfies the initial condition $\rho_i(0)=\text{Tr}_{k\neq i}\{\rho(0)\}$.
The symbol $\text{Tr}_{k \neq i}$ means tracing with respect to ``the other spin''.
Equation \eqref{Reduce Eq. of Motion} clearly shows that in correspondence to each $\rho(t)$ such that $[H_I(t),\rho(t)]=0$ at any time instant, the $i$-th reduced density operator $\rho_i(t)$ satisfies the following Cauchy problem
\begin{equation}\label{Reduced Motion Eq.}
i\dot{\rho}_i(t) = [H_i,\rho_i(t)], \quad \rho_i(0)=\text{Tr}_{k \neq i}\{\rho(0)\}.
\end{equation}
Since $H$ and $H_I$ commute with both $\hat{\mathbf{J}}^2$ and $\hat{J}^z$, it is easy to convince oneself that any density matrix $\rho(0)$, at any time satisfying $[H_I(t),\rho(t)]=0$, may be represented in the coupled basis as
\begin{equation}\label{IFE states}
[\rho(0)]_{CB}=\bigoplus_{j=|j_1-j_2|}^{j_1+j_2} \rho^{(j)}
\end{equation}
$\rho^{(j)}$ being a $(2j+1)$-dimensional semi-positive definite matrix such that $\rho(0)$ is a density matrix.
As a consequence, when $\rho(0)$ belongs to the class of initial conditions given by Eq. \eqref{IFE states}, the solution of the Cauchy problem \eqref{Reduced Motion Eq.} may be written down as follows 
\begin{equation}
\rho_i(t)=U_{i}(t) \rho_i(0) U_{i}^\dagger(t),
\end{equation}
where $U_{i}(t)$ is the unitary operator governing the SU(2) time evolution of the spin $\hat{\textbf{j}}_i$ when $\lambda = 0$.

In words, the symmetries of the Hamiltonian, under the condition \eqref{IFE states}, guarantee that each spin subsystem evolves as if the other one were absent, that is undergoing no influence stemming from the coupling term.
It is worthwhile to remark that such a property holds whatever the magnetic field time-dependence is.
At the light of this result we understand better and more deeply the result in Eq. \eqref{P jj -j-j}: since the factorized initial state $\ket{j_1,j_2}$ of the compound system belongs to the class of initial conditions given in Eq. \eqref{IFE states}, then, the joint probability of finding the two spins in the state $\ket{-j_1,-j_2}$ is nothing but the probability $|b|^{4j_1}$ of finding the spin $j_1$ in its state $\ket{-j_1}$ multiplied by the probability $|b|^{4j_2}$ of finding the spin $j_2$ in its state $\ket{-j_2}$.

In addition we recognize that the class of initial states given by Eq. \eqref{IFE states} collects states being Interaction Free Evolving (IFE) states, recently reported in literature \cite{IFE1,IFE2,IFE3}.
By definition they are pure or mixed states of a binary system evolving in time as if the interaction between the two subsystems were absent.
We point out that, of course, any generic initial condition of the compound system presenting coherence terms between the different dynamically invariant subspaces of $H$ and $\hat{\mathbf{J}}^2$ does not manifest such a peculiar dynamical feature since the reduced dynamics of the two spins is influenced by the existing isotropic Heisenberg coupling between the two spins.

It is finally worthwhile to observe that the initial entanglement between the two spins $\hat{\mathbf{j}}_1$ and $\hat{\mathbf{j}}_2$ in an arbitrary state of the class singled out by Eq. \eqref{IFE states}, does not change in time, whatever the entanglement measure adopted is.
The physical reason may be traced back to the dynamical quenching of the interaction term stemming, in turn, from constraints on the evolution imposed by the symmetry properties possessed by our Hamiltonian model \eqref{Hamiltonian}.


\section{Three Exemplary cases: \hspace{3cm} $j_1=j_2=1/2$; $j_1=2j_2=1$; $j_1=j_2=1$}\label{Ex Cases}

In this section we apply our general procedure explained in the first section to three exemplary cases:
$j_1=j_2=1/2$; $j_1=1, j_2=1/2$; $j_1=j_2=1$.
In the first case we retrace the general procedure explained in the first section, while for two last cases we give only the final interesting result.

Let us consider the case $j_1=j_2=1/2$ so that
\begin{equation}
\hat{\mathbf{j}}_i = {1\over 2} \hat{\bm{\sigma}}_i, \quad i=1,2,
\end{equation}
$\hat{\bm{\sigma}}_i \equiv (\hat{\sigma}_i^x,\hat{\sigma}_i^y,\hat{\sigma}_i^z)$ being the Pauli vector of $i$-th spin.
In this instance, the corresponding Hamiltonian model reads as
\begin{equation}\label{Hamiltonian spin 1/2s}
H ={\omega_x(t) \over 2} \hat{\Sigma}^x+{\omega_y(t) \over 2} \hat{\Sigma}^y+{\Omega(t) \over 2} \hat{\Sigma}^z - \dfrac{\lambda}{4} \hat{\bm{\sigma}}_1 \cdot \hat{\bm{\sigma}}_2
\end{equation}
with $[\omega_x(t),\omega_y(t),\Omega(t)]\equiv\bm{\Omega}(t)$ and $\hat{\Sigma}^\alpha = \hat{\sigma}_1^\alpha+\hat{\sigma}_2^\alpha$ $(\alpha=x,y,z)$.

As explained in the previous general section, the conservation of $\mathbf{\hat{J}}^2$ implies the existence of two orthogonal, dynamically invariant subspaces $\mathcal{H}^{(0)}$ and $\mathcal{H}^{(1)}$ such that ($\mathcal{H}$ denotes the total four dimensional Hilbert space of the two spin 1/2's)
\begin{equation}
\mathcal{H}=\mathcal{H}^{(0)} \oplus \mathcal{H}^{(1)}
\end{equation}
and the two subspaces are spanned respectively by 
\begin{equation}\label{S^2=0 eigenstate}
 \ket{j=0,m=0}={\ket{+-}-\ket{-+} \over \sqrt{2}} \equiv \ket{\Psi^-},
\end{equation}
and
\begin{equation}\label{S^2=1 eigenstates}
\begin{aligned}
& \ket{j=1,m=1}\equiv\ket{++}, \quad \ket{j=1,m=-1}\equiv\ket{--}, \\
&\ket{j=1,m=0}={\ket{+-}+\ket{-+} \over \sqrt{2}}\equiv\ket{\Psi^+}.
\end{aligned}
\end{equation}
The four states $\{ \ket{++},\ket{+-},\ket{-+},\ket{--} \}$, appearing in Eqs. \eqref{S^2=0 eigenstate} and \eqref{S^2=1 eigenstates}, are the four orthonormalized standard factorized eigenstates of $\hat{\Sigma}^z$ and is assumed as the standard basis of $\mathcal{H}$.

By Eq. \eqref{H direct sum}, the representation $(H)_{CB}$ of $H$ in the coupled basis ordered as follows $\{ \ket{++},\ket{\Psi^+},\ket{--},\ket{\Psi^-} \}$, may be written down as
\begin{equation}\label{H tilde}
(H)_{CB}=H^{(1)} \oplus H^{(0)}
\end{equation}
with
\begin{equation}
\begin{aligned}
H^{(1)}=
\left(
\begin{array}{cccc}
 \Omega-{\lambda \over 4}  & \frac{\omega }{\sqrt{2}} & 0 \\
 \frac{\text{$\omega^{*} $}}{\sqrt{2}} & -{\lambda \over 4} & \frac{\omega }{\sqrt{2}} \\
 0 & \frac{\text{$\omega^{*} $}}{\sqrt{2}} & -\Omega-{\lambda \over 4} \\
\end{array}
\right),
\quad
H^{(0)}={3 \over 4}\lambda,
\end{aligned}
\end{equation}
where $\omega = \omega_x - i \omega_y$.
We see, as expected, that $H^{(1)}$ is (up to the constant term $-{\lambda \over 4}$) the SU(2) Hamiltonian of a fictitious spin 1 subjected to the same common magnetic field $\mathbf{B}(t)$ acting upon the two spin 1/2's of our model \eqref{Hamiltonian spin 1/2s}.
To this end it is enough to map the three states $\{ \ket{++},\ket{\Psi^+},\ket{--} \}$ into the eigenstates of the $z$-component of the fictitious spin 1 $\{ \ket{1},\ket{0},\ket{-1} \}$.
It is important to underline that this result agrees with that brought to light in \cite{Bagrov1}.

Such an identification is of relevance since, as explained before, it is well known that the correspondent evolution operator may be immediately written down after determining the evolution operator of an auxiliary spin 1/2 subjected to the same effective time-dependent magnetic field $\bm{\Omega}(t)$ \cite{Hioe}.
As a consequence, according to Eq. \eqref{U direct sum}, the matrix representation $(U)_{CB}$ in the coupled basis of our physical system of the evolution operator $U$ related to $H$ may be cast as follows
\begin{equation}\label{P1 U P1}
\begin{aligned}
(U)_{CB}&=U^{(1)} \oplus U^{(0)}= \\
&=e^{i{\lambda \over 4}t}
\begin{pmatrix}
a^{2} & \sqrt{2}ab & b^{2} \\
-\sqrt{2}ab^{*} & |a|^2-|b|^2 & \sqrt{2}a^{*}b \\
{b^{*}}^{2} & -\sqrt{2}a^{*}b^{*} & {a^{*}}^{2}
\end{pmatrix}
\oplus e^{-i{3\lambda \over 4}t}
\end{aligned}
\end{equation}
where $a=a(t)$ and $b=b(t)$ satisfy $|a|^2+|b|^2=1$ at any time.
The unitary time-dependent matrix governing the quantum dynamics of the auxiliary spin 1/2 mentioned above is defined in Eq. \eqref{TEO spin 1/2}.
We emphasize that in the context of the two spin 1/2's problem such a bidimensional evolution operator governs the quantum dynamics of each spin 1/2 in our model if $\lambda=0$.

Observing, now, that the unitary matrix $T$ accomplishing the transformation from the ordered coupled basis into the standard basis is
\begin{equation}\label{CB matrix spin 1/2's}
T=
\left(
\begin{array}{cccc}
 1 & 0 & 0 & 0 \\
 0 & \frac{1}{\sqrt{2}} & 0 & \frac{1}{\sqrt{2}} \\
 0 & \frac{1}{\sqrt{2}} & 0 & -\frac{1}{\sqrt{2}} \\
 0 & 0 & 1 & 0
\end{array}
\right),
\end{equation}
it is easy to convince oneself that the matrix representation $(U)_{SB}$ of the  evolution operator $U$ of our system in the standard basis, ordered as $\{ \ket{++}, \ket{+-}, \ket{-+}, \ket{--} \}$, assumes the following form
\begin{widetext}
\begin{equation}\label{TEO H}
T (U)_{CB} T^\dagger = (U)_{SB} = e^{i{\lambda \over 4}t}
\left(
\begin{array}{cccc}
 a^2 & a b & a b & b^2 \\
 -a b^* & \frac{1}{2} (e^{-i{\lambda}t} + |a|^2- |b|^2) & \frac{1}{2} (-e^{-i{\lambda}t} + |a|^2- |b|^2) & a^* b \\
 -a b^* & \frac{1}{2} (-e^{-i{\lambda}t} + |a|^2- |b|^2) & \frac{1}{2} (e^{-i{\lambda}t} + |a|^2- |b|^2) & a^* b \\
 (b^*)^2 & -a^* b^* & -a^* b^* & (a^*)^2
\end{array}
\right).
\end{equation}

Analogously, exploiting the same arguments and following the same procedure and approach, for the case $j_1=1$ and $j_2=1/2$ we get
(the standard basis is ordered as $\{ \ket{1,+},\ket{1,-},\ket{0,+},\ket{0,-},\ket{-1,+},\ket{-1,-} \}$)
\begin{footnotesize}
\begin{equation}
\begin{aligned}
&(U)_{SB}=e^{i{\lambda \over 2}t} \times \\ \times
&\left(
\begin{array}{cccccc}
 a^3 & a^2 b & \sqrt{2} a^2 b & \sqrt{2} a b^2 & a b^2 & b^3 \\
 -a^2 b^* & \frac{1}{3} a \left(|a|^2-2 |b|^2+2 e^{-\frac{3 \lambda t }{2}}\right) & \frac{\sqrt{2}}{3} a \left(|a|^2-2 |b|^2-e^{-\frac{3 \lambda t }{2}}\right) & \frac{\sqrt{2}}{3} b \left(2 |a|^2-|b|^2+e^{-\frac{3 \lambda t }{2}}\right) & \frac{1}{3} b \left(2 |a|^2-|b|^2-2 e^{-\frac{3 \lambda t }{2}}\right) & a^* b^2 \\
 -\sqrt{2} a^2 b^* & \frac{\sqrt{2}}{3} a \left(|a|^2-2 |b|^2-e^{-\frac{3 \lambda t }{2}}\right) & \frac{1}{3} a \left(2 |a|^2-4 |b|^2+e^{-\frac{3 \lambda t }{2}}\right) & \frac{1}{3} b \left(4 |a|^2-2 |b|^2-e^{-\frac{3 \lambda t }{2}}\right) & \frac{\sqrt{2}}{3} b \left(2 |a|^2-|b|^2+e^{-\frac{3 \lambda t }{2}}\right) & \sqrt{2} a^* b^2 \\
 \sqrt{2} a (b^*)^2 & \frac{\sqrt{2}}{3} b^* \left(-2 |a|^2+|b|^2-e^{-\frac{3 \lambda t }{2}}\right) & \frac{1}{3} b^* \left(-4 |a|^2+2 |b|^2+e^{-\frac{3 \lambda t }{2}}\right) & \frac{1}{3} a^* \left(2 |a|^2-4 |b|^2+e^{-\frac{3 \lambda t }{2}}\right) & \frac{\sqrt{2}}{3}  a^* \left(|a|^2-2 |b|^2-e^{-\frac{3 \lambda t }{2}}\right) & \sqrt{2} (a^*)^2 b \\
 a (b^*)^2 & \frac{1}{3} b^* \left(-2 |a|^2+|b|^2+2 e^{-\frac{3\lambda t  }{2}}\right) & \frac{\sqrt{2}}{3} b^* \left(-2 |a|^2+|b|^2-e^{-\frac{3 \lambda t }{2}}\right) & \frac{\sqrt{2}}{3} a^* \left(|a|^2-2 |b|^2-e^{-\frac{3 \lambda t }{2}}\right) & \frac{1}{3} a^* \left(|a|^2-2 b |b|^2+2 e^{-\frac{3 \lambda t }{2}}\right) & (a^*)^2 b \\
 -(b^*)^3 & a^* (b^*)^2 & \sqrt{2} a^* (b^*)^2 & -\sqrt{2} (a^*)^2 (b^*) & -(a^*)^2 b^* & (a^*)^3
\end{array}
\right).
\end{aligned}
\end{equation}
\end{footnotesize}
In this case the unitary transformation matrix $W$ to get $(U)_{SB}$ from $(U)_{CB}$, namely $(U)_{SB}=W(U)_{SB}W^\dagger$, reads
\begin{equation}
W=
\left(
\begin{array}{cccccc}
 1 & 0 & 0 & 0 & 0 & 0 \\
 0 & \frac{1}{\sqrt{3}} & 0 & 0 & \sqrt{\frac{2}{3}} & 0 \\
 0 & \sqrt{\frac{2}{3}} & 0 & 0 & -\frac{1}{\sqrt{3}} & 0 \\
 0 & 0 & \sqrt{\frac{2}{3}} & 0 & 0 & \frac{1}{\sqrt{3}} \\
 0 & 0 & \frac{1}{\sqrt{3}} & 0 & 0 & -\sqrt{\frac{2}{3}} \\
 0 & 0 & 0 & 1 & 0 & 0
\end{array}
\right).
\end{equation} 
For two spin 1's the matrix representation $(U)_{SB}$ of time evolution operator $U$ in the standard basis can be got analogously by $(U)_{SB}=C(U)_{CB}C^\dagger$.
$C$, being the unitary Clebsh-Gordan matrix for two spin 1's accomplishing the transformation from the ordered coupled basis into the standard one, 
reads
\begin{equation}\label{CB matrix spin 1's}
C=
\left(
\begin{array}{ccccccccc}
 1 & 0 & 0 & 0 & 0 & 0 & 0 & 0 & 0 \\
 0 & \frac{1}{\sqrt{2}} & 0 & 0 & 0 & \frac{1}{\sqrt{2}} & 0 & 0 & 0 \\
 0 & 0 & \frac{1}{\sqrt{6}} & 0 & 0 & 0 & \frac{1}{\sqrt{2}} & 0 & \frac{1}{\sqrt{3}} \\
 0 & \frac{1}{\sqrt{2}} & 0 & 0 & 0 & -\frac{1}{\sqrt{2}} & 0 & 0 & 0 \\
 0 & 0 & \sqrt{\frac{2}{3}} & 0 & 0 & 0 & 0 & 0 & -\frac{1}{\sqrt{3}} \\
 0 & 0 & 0 & \frac{1}{\sqrt{2}} & 0 & 0 & 0 & \frac{1}{\sqrt{2}} & 0 \\
 0 & 0 & \frac{1}{\sqrt{6}} & 0 & 0 & 0 & -\frac{1}{\sqrt{2}} & 0 & \frac{1}{\sqrt{3}} \\
 0 & 0 & 0 & \frac{1}{\sqrt{2}} & 0 & 0 & 0 & -\frac{1}{\sqrt{2}} & 0 \\
 0 & 0 & 0 & 0 & 1 & 0 & 0 & 0 & 0
\end{array}
\right)
\end{equation}
and $(U)_{CB}$, the matrix representation of $U$ in the coupled basis, is given by
\begin{equation}\label{U spin 1's}
  (U)_{CB}= e^{i\lambda t}U^{(2)} \oplus e^{-i\lambda t}U^{(1)} \oplus e^{-2i\lambda t}U^{(0)},
\end{equation}
with
\begin{equation}
\begin{aligned}
&U^{(0)}=1,
\qquad
U^{(1)}=
\begin{pmatrix}
a^{2} & \sqrt{2}ab & b^{2} \\
-\sqrt{2}ab^{*} & |a|^2-|b|^2 & \sqrt{2}a^{*}b \\
{b^{*}}^{2} & -\sqrt{2}a^{*}b^{*} & {a^{*}}^{2}
\end{pmatrix},
\\
&U^{(2)}=
\begin{pmatrix}
    a^4 & 2a^3b &\sqrt{6}a^2b^2 & 2ab^3 & b^4\\
    -2a^3b^* & (|a|^2-3|b|^2)a^2 & \sqrt{6}(|a|^2-|b|^2)ab &
    (3|a|^2-|b|^2)b^2 & 2a^*b^3\\
   \sqrt{6}a^2(b^*)^2 & -\sqrt{6}(|a|^2-|b|^2)ab^* &
   1-6|a|^2|b|^2 & \sqrt{6}(|a|^2-|b|^2)a^*b &
   \sqrt{6}(a^*)^2b^2\\
    -2a(b^*)^3 & (3|a|^2-|b|^2)(b^*)^2 &
    -\sqrt{6}(|a|^2-|b|^2)a^*b^* &
    (|a|^2-3|b|^2)(a^*)^2 & 2(a^*)^3b\\
   (b^*)^4 & -2a^*(b^*)^3 &\sqrt{6}(a^*)^2(b^*)^2 & -2(a^*)^3b^* & (a^*)^4
  \end{pmatrix}.
  \end{aligned}
\end{equation}

\end{widetext}

\section{Highlighting dynamical effects due to $H_I$}\label{Int Evid}

To bring to light effects witnessing peculiar features in the quantum dynamic of the two coupled GRSs, we have to consider initial states generating coherences between different dynamically invariant subspaces of $H$. 
To this end, let us consider the $(2j_1+1)(2j_2+1)$ factorized states of the standard basis $\{ \ket{m_1,m_2}; -j_1 \leq m_1 \leq j_1, -j_2 \leq m_2 \leq j_2 \}$, ordered as 
\begin{equation}
\begin{aligned}
\Bigl\{ \ket{j_1,j_2}\equiv\ket{e_1}, \ket{j_1,j_2-1}\equiv\ket{e_2}, \dots, \ket{j_1,-j_2}\equiv\ket{e_{2j_2+1}}, \\
\ket{j_1-1,j_2}\equiv\ket{e_{2j_2+2}}, \ket{j_1-1,j_2-1}\equiv\ket{e_{2j_2+3}}, \dots, \\
\ket{-j_1,j_2}\equiv\ket{e_{2j_2(2j_1+1)}}, \dots, \ket{-j_1,-j_2}\equiv\ket{e_{(2j_1+1)(2j_2+1)}} \Bigr\},
\end{aligned}
\end{equation}
The projections of the factorized state $\ket{\psi(0)}=\ket{j_1,j_2-1}$ in the two invariant subspaces of $\hat{\mathbf{J}}^2$, labelled by $(j_1+j_2)$ and $(j_1+j_2-1)$, do not vanish and then the evolution of $\ket{\psi(0)}$ may be expressed as 
\begin{equation}\label{+- of t}
\ket{\psi(t)}=\sum_k U_{k2} \ket{e_k}
\end{equation}
where $k$ runs from 1 to $(2j_1+1)(2j_2+1)$ generating the entries $U_{k2}$ in the second column of $(U)_{SB}$, in accordance with our ordered standard basis.

To reach our goal, it is also important to choose appropriately the physical observable to be investigated.
If we consider, e.g., the third component of the total spin angular momentum of the system $\hat{J}^z=\hat{j}_1^z+\hat{j}_2^z$, it is easy to verify that it commutes with the isotropic Heisenberg interaction $H_I=-\lambda \hat{\mathbf{j}}_1 \cdot \hat{\mathbf{j}}_2$.
Since, in addition, $[H_0,H_I]=0$ at any time instant, then
\begin{equation}
\begin{aligned}
&\average{\psi(t)|\hat{J}^z|\psi(t)}=\average{\psi_0(t)|\hat{J}^z|\psi_0(t)}= \\
&=\average{\psi_{01}(t)|\hat{j}_1^z|\psi_{01}(t)}+\average{\psi_{02}(t)|\hat{j}_2^z|\psi_{02}(t)}
\end{aligned}
\end{equation}
where $\ket{\psi_0(t)}=U_0(t)\ket{\psi(0)}$, $\ket{\psi_{01}(t)}=U_{01}\ket{j_1}$ and $\ket{\psi_{02}(t)}=U_{02}\ket{j_2-1}$, with $\hat{j}_1^z\ket{j_1}=j_1\ket{j_1}$ and $\hat{j}_2^z\ket{j_2-1}=(j_2-1)\ket{j_2-1}$.
In these expressions $U_0(t)$ is the unitary evolution operator generated by $H_0(t)$, whereas $U_{0 i}(t)$ is that generated by $\bm{\Omega}\cdot\hat{\mathbf{j}}_i$.
Thus, we predict the independence of $\average{\psi(t)|\hat{J}^z|\psi(t)}$ from $\lambda$, regardless of the specific magnetic field acting upon the two-spin system.
It is indeed possible to convince oneself that
\begin{eqnarray}\label{Jz of t +-}
\average{\hat{J}^z(t)} = 0, \quad &&\text{if $j_1=j_2=1/2$}  \\
\average{\hat{J}^z(t)} = {|a|^2-|b|^2 \over 2}, \quad &&\text{if $j_1=2j_2=1$}  \\
\average{\hat{J}^z(t)} = |a|^2-|b|^2, \quad &&\text{if $j_1=j_2=1$},
\end{eqnarray}

In order to predict a visible effect of the coupling between the spins, we calculate the time-dependence of the mean value of $\hat{j}_1^z$, getting
\begin{equation}
\begin{aligned}
\average{\psi(t)|\hat{j}_1^z|\psi(t)}=\sum_{i=0}^{2j_1} \hspace{0,2cm} \sum_{k=1+i(2j_2+1)}^{(i+1)(2j_2+1)} (j_1-i) |U_{k2}|^2,
\end{aligned}
\end{equation}
which in the three particular cases under scrutiny, leads respectively to the following explicit expressions
\begin{eqnarray}
\average{\hat{j}_1^z(t)}=&&{1\over 2}(|a|^2-|b|^2)\cos({\lambda}t), \label{J1z +-t 1/2-1/2} \\
 \average{\hat{j}_1^z(t)}=&&{1\over 9}(|a|^2-|b|^2)\Bigl[5+4\cos({3\lambda \over 2}t)\Bigr], \label{J1z +-t 1-1/2} \\
 \average{\hat{j}_1^z(t)}=&&{1\over 2}\Bigl[|a|^2-|b|^2+\nonumber \\ 
 &&+\Bigl(1-2|a|^2|b|^2(|a|^2-2|b|^2)\Bigr)\cos({2\lambda}t)\Bigr], \label{J1z +-t 1-1}
\end{eqnarray}
with the short notation $\average{\hat{j}_1^z(t)}=\average{\psi(t)|\hat{j}_1^z|\psi(t)}$.

It is remarkable that by measuring the magnetization time-dependence of any one of the two spin subsystems we may experimentally recover information about the coupling strength, regardless of the applied magnetic field.
This fact, in view of Eqs. \eqref{J1z +-t 1/2-1/2}, \eqref{J1z +-t 1-1/2} and \eqref{J1z +-t 1-1}, enables us to check experimentally whether a direct interaction between the two spins exists or at least plays a non-negligible role in the Hamiltonian model describing the two-spin system in a given physical scenario.

\Ignore{
We conclude this section by pointing out an interesting feature of the entanglement established between the two spins $\hat{\mathbf{j}}_1$ and $\hat{\mathbf{j}}_2$, discussing for simplicity the case $j_1=j_2=1/2$.
Since any initial eigenstate of $(\hat{\mathbf{j}}_1+\hat{\mathbf{j}}_2)^2$ evolves keeping constant the level of entanglement between the two spins, we prepare the system in the initial state 
\begin{equation}\label{Class Bell states}
\ket{\psi}=\cos(\theta)\ket{\Psi^+}+e^{i\phi}\sin(\theta)\ket{\Psi^-}
\end{equation}
and evaluate the time evolution of the entanglement established between the two spins using the concurrence $C(t)$ \cite{Wootters-Hill,Wootters} or equivalently the negativity $\mathcal{N}(t)$ \cite{Vidal-Werner,Grudka} getting
\begin{equation}
C(t)=\mathcal{N}(t)=\bigl|\cos^2(\theta)-e^{i2(\phi-\lambda t)}\sin^2(\theta)\bigr|.
\end{equation}
The point deserving to be emphasized is that $C(t)$ does not depend on the applied magnetic field but only on the initial state parameters $\theta$ and $\phi$ as well as on the coupling strength $\lambda$.
We stress that such a property is peculiar of the case $j_1=j_2=1/2$ under scrutiny, in the sense that in the general case a dependence of $C(t)$ on the applied magnetic field, indeed, appears, since this time its expression includes $a(t)$ and $b(t)$. 
}

\section{Noisy Magnetic Field}\label{Noise}

In this section we wish to take into consideration the possibility that our two-spin system is subjected to a random magnetic field acting together with the external controllable one.
The origin of such a random field, in the case of pair of interacting nanomagnets hosted in a solid matrix, might for example be the nuclear spin environment around the two-spin system \cite{Pokrovsky2}.
In other physical scenarios, however, noise might be traced back to different physical mechanisms (see Ref. \cite{Kenmoe2} and references therein).

For simplicity we take the noise as a classical random field $\mathbf{B}_{r}(t)$ and introduce $\bm{\eta}(t)=\sum_{i=x,y,z} \eta_i(t) \mathbf{c}_i\equiv-\gamma\mathbf{B}_{r}(t)$ in analogy with $\bm{\Omega}(t)$.
In this section we adopt a model where the two spins are equal and with the same $\gamma$.
As in Ref. \cite{Pokrovsky1}, the random vector $\bm{\eta}(t)$, with Gaussian realizations, is supposed to be characterized by the following general correlation tensor
\begin{equation}\label{Noise correlator}
\average{\eta_i(t) \eta_j(t')}=g_{ij}(\Lambda |t-t'|),
\end{equation}
$\Lambda$ being the inverse characteristic decay time of the correlators $g_{ij}$ which in turn are assumed of the same order of magnitude.
Equation \eqref{Noise correlator} defines a fast colored noise when $\Lambda \rightarrow \infty$, which reproduces a white noise under the further condition that $\average{\eta_i(t) \eta_j(t')}$ turns into a delta function \cite{Pokrovsky1}.

The Hamiltonian describing our two equal, distinguishable spin systems in the presence of such a source of colored noise may be cast in the following form
\begin{equation}\label{Ham with noise}
H(t)=\sum_{i=1}^2 [\bm{\Omega}(t)+\bm{\eta}(t)] \cdot \hat{\textbf{j}}_i-\lambda \hat{\textbf{j}}_1 \cdot \hat{\textbf{j}}_2.
\end{equation}
The assumption of a random magnetic field homogeneously acting upon both spins is evident in Eq. \eqref{Ham with noise}.
We stress that it has been adopted in recent literature relevant to the problem under scrutiny \cite{Das Sarma PRB, Das Sarma PRA}.
 
Comparing the symmetries of this model with those of our original model given in Eq. \eqref{Hamiltonian}, we claim that each invariant subspace of ${\hat{\mathbf{J}}}^2=({\hat{\mathbf{j}}}_1+{\hat{\mathbf{j}}}_2)^2$ is a dynamically invariant subspace of $H(t)$ as well, even in presence of a fluctuating magnetic field.
Noteworthy differences in the quantum dynamics of the two-spin system however emerge as a direct consequence of the noisy source.

To appreciate this point, consider, for instance, the time evolution of the initial factorized state $\ket{j_1,j_2}$.
In the previous sections we have demonstrated that, when the fluctuating field is absent, this state has an IFE nature, so that the probability $P_{j_1,j_2}^{-j_1,-j_2}(t)$ of finding the two spins in $\ket{-j_1,-j_2}$ at a generic time instant is nicely expressible as a product of the relevant probabilities, $P_{j_1}^{-j_1}(t)$ and $P_{j_2}^{-j_2}(t)$ governing the transition paths of the two spins.
Such a property, still true for any specific realization of the fluctuating field, is lost when we average over all the possible realizations of this field.
In other words, in the presence of a random magnetic field $P_{j_1,j_2}^{-j_1,-j_2}(t) \neq P_{j_1}^{-j_1}(t) P_{j_2}^{-j_2}(t)$, since now all these three involved probabilities result from the resolution of the relevant dynamical equations and from the average processes.

However, the quantum evolution of $\ket{j_1,j_2}$ generated by the Hamiltonian model \eqref{Ham with noise} may be mapped into that experienced by a single fictitious spin $\hat{\mathbf{j}}$, of maximum projection $j_1+j_2$, under the Hamiltonian given by Eq. \eqref{Hj} where the superscript $j$ and the field $\bm{\Omega}(t)$ are respectively replaced by $j_1+j_2$ and $\bm{\Omega}(t)+\bm{\eta}(t)$, in accordance with Eq. \eqref{Ham with noise}.  
Thus, it is legitimate to identify $P_{j_1+j_2}^{-(j_1+j_2)}(t)$, that is the probability of a complete inversion of the fictitious spin $\hat{\mathbf{j}}$ at a generic time instant $t$, with $P_{j_1,j_2}^{-j_1,-j_2}(t)$ denoting the joint inversion probability of both $\hat{\mathbf{j}}_1$ and $\hat{\mathbf{j}}_2$.
We stress this equality, namely
\begin{equation}\label{Joint general LZ prob noise}
P_{j_1,j_2}^{-j_1,-j_2}(t)=P_{j_1+j_2}^{-(j_1+j_2)}(t)
\end{equation}
holds whatever the fields $\bm{\Omega}(t)$ and $\bm{\eta}(t)$ are.

In order to show clearly the validity and the usefulness of our approach also in this instance, in the following we are going to concentrate on a Gaussian process described by a classical fast transverse ($\eta_z=0$) noise field characterized by the general time correlation functions in Eq. \eqref{Noise correlator}.
Our scope is to bring to light its influence on joint Landau-Zener transitions of two identical distinguishable spins whose dynamics is ruled out by the Hamiltonian \eqref{Ham with noise}, accordingly specialized by putting $\bm{\Omega}(t)\equiv (\Delta,0,\alpha t)$.

The physical interest towards this specific experimental scenario stems from the fact that adiabatic LZ transitions might play an applicative role in quantum computing \cite{Bell}.
To address such an implementation it then becomes crucial to be aware of effects traceable back to the unavoidable presence of noise. 

In view of a comparison between the ideal scenario and the one where the additional fluctuating field too is taken into account, it is useful to start by reporting explicit expressions of the relevant LZ transition probabilities.
To this end we observe that
\begin{equation}
P_{j_1,j_2}^{-j_1,-j_2}(t=\infty) = P_{j_1}^{-j_1}(\infty)P_{j_2}^{-j_2}(\infty) = \bigl[ P_{1/2}^{-1/2}(\infty) \bigr]^{2(j_1+j_2)}
\end{equation}
where the last SU(2)-based equality follows from Eq. \eqref{P jj -j-j}.
The transition probability $P_{1/2}^{-1/2}(\infty)$ coincides of course with the celebrated LZ transition probability $P_{LZ}$, though derived independently by different people in the same year \cite{Landau,Zener,Stuckelberg,Majorana}, so that, we get
\begin{equation}\label{Joint general LZ prob}
P_{j_1,j_2}^{-j_1,-j_2}(t=\infty) = \bigl[ 1-e^{-2\pi\Gamma} \bigr]^{2(j_1+j_2)},
\end{equation}
where $\Gamma={\Delta^2 \over 4\alpha}$ is the Landau-Zener parameter.

It is of relevance to point out that Eq. (60) of Ref. \cite{Pokrovsky2} furnishes the explicit expression of $P_{j_1+j_2}^{-(j_1+j_2)}(t=\infty)$ when a transverse random field is present too and then, in view of Eq. \eqref{Joint general LZ prob noise}, we may write the solution of the joint LZ transitions of two interacting spins affected by noise.
Here, for simplicity, we confine ourselves to the case $j_1=j_2=1/2$, where, in view of the analysis of this section, searching joint LZ transition probability $P_{1/2,1/2}^{-1/2,-1/2} \equiv P_{++}^{--}$ amounts at investigating the LZ effect in a symmetric effective three-level system.

The asymptotic ($t=\infty$) populations for a three-level system subjected to a noisy LZ scenario as given in Refs. \cite{Pokrovsky2,Kenmoe1}, may be directly reinterpreted in terms of the two-spin-1/2 system language, yielding
\begin{widetext}
\begin{eqnarray}
&&P_{++}^{--}={1 \over 3}\Bigl[1+({3 \over 2}e^{-\theta/2}+{1 \over 2}e^{-3\theta/2})(1-4e^{-2\pi\Gamma}+3e^{-4\pi\Gamma})-
({3 \over 2}e^{-\theta/2}-{1 \over 2}e^{-3\theta/2})(3e^{-4\pi\Gamma}-2e^{-2\pi\Gamma})\Bigr], \label{Prob noise ++--}
\end{eqnarray}
\begin{eqnarray}
&&P_{++}^{\Psi^+}=2P_{++}^{+-}=2P_{++}^{-+}={1 \over 3}\Bigl[1-e^{-3\theta/2}(1-4e^{-2\pi\Gamma}+3e^{-4\pi\Gamma})-
e^{-3\theta/2}(3e^{-4\pi\Gamma}-2e^{-2\pi\Gamma})\Bigr], \\
&&P_{++}^{++}={1 \over 3}\Bigl[1-({3 \over 2}e^{-\theta/2}-{1 \over 2}e^{-3\theta/2})(1-4e^{-2\pi\Gamma}+3e^{-4\pi\Gamma})+
({3 \over 2}e^{-\theta/2}+{1 \over 2}e^{-3\theta/2})(3e^{-4\pi\Gamma}-2e^{-2\pi\Gamma})\Bigr], \label{Prob noise ++++}
\end{eqnarray}
of course fulfilling the normalization condition $P_{++}^{++}+P_{++}^{\Psi^+}+P_{++}^{--}=1$.
These three probabilities are notationally of the form: $P_{\phi}^{\phi'}$ expressing the probability that the compound system evolves from the initial state $\ket{\phi}$ to the final state $\ket{\phi'}$ of the two-spin system.
For example $P_{++}^{\Psi^+}$ denotes the transition probability from the state $\ket{++}\equiv\ket{1/2,1/2}$ to the state $\ket{\Psi^+}$ defined in Eq. \eqref{S^2=1 eigenstates}.
Moreover, $\theta={4\pi \over \alpha}\mathcal{R}(0)$ with $\mathcal{R}(0)=\average{\eta_x(t)\eta_x(t')}\bigl|_{t=t'}+\average{\eta_y(t)\eta_y(t')}\bigl|_{t=t'}=g_{xx}^2(0)+g_{yy}^2(0)$, in view of Eq.\eqref{Noise correlator}.
It is immediate to check that putting $\theta=0$ in Eq. \eqref{Prob noise ++--} we recover Eq. \eqref{Joint general LZ prob} where $j_1+j_2=1$ has been used.
\end{widetext}

The joint LZ transition probability $P_{++}^{--}$ is plotted in Fig. 2 as a function of $\Gamma$ treating $\theta$ as a parameter and \textit{vice versa} in Fig. 3.
\begin{figure}[tbph]
\centering
{\includegraphics[width=\columnwidth]{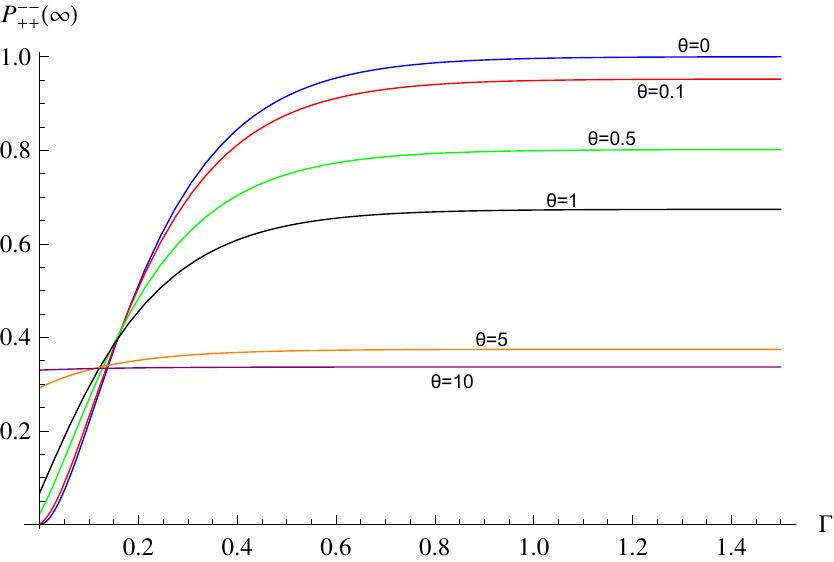} }
\caption{Plot of $P_{++}^{--}(\infty)$ as a function of $\Gamma$ for different values of $\theta$.}
\end{figure}
\begin{figure}[tbph]
\centering
{\includegraphics[width=\columnwidth]{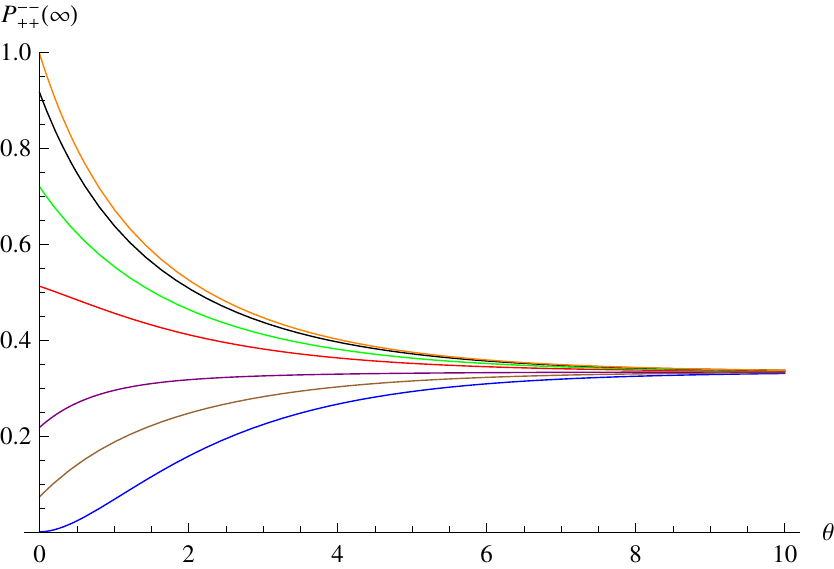} }
\caption{Plot of $P_{++}^{--}(\infty)$ as a function of $\theta$ for different values of $\Gamma$ (from top to bottom): $\Gamma=1$, $\Gamma=0.5$, $\Gamma=0.3$, $\Gamma=0.2$, $\Gamma=0.1$, $\Gamma=0.05$, $\Gamma=0$.}
\end{figure}
Both plots show that at very big noise the joint LZ transition probability becomes more and more insensitive to the external applied transverse magnetic field sharing its value 1/3 with the other two populations $P_{++}^{\Psi^+}$ and $P_{++}^{++}$.
Moreover, from the previous plots we understand that the effect of the noise is different depending on the value of the controllable applied magnetic field.
Indeed, it is possible to verify that for
\begin{equation}
0\leq\Gamma\leq{1 \over 2\pi}\log\Bigl( {3+\sqrt{3} \over 2} \Bigr)
\end{equation}
$P_{++}^{--}$ is always favoured by the random magnetic field, while for
\begin{equation}
\Gamma\geq{1\over 2\pi}\log(3)
\end{equation}
$P_{++}^{--}$ is always countered by the noisy field.
In the $\Gamma$-interval
\begin{equation}\label{Last Range Gamma}
{1 \over 2\pi}\log\Bigl( {3+\sqrt{3} \over 2} \Bigr)<\Gamma<{1\over 2\pi}\log(3),
\end{equation}
we instead have two ranges of values of $\theta$ in which the random field acts first favouring and then hindering the transition as it can be seen by Fig. 4.
In other words our analysis unveils the loss of the monotonicity of $P_{++}^{--}$ versus $\theta$, clearly shown in Fig. 4, when $\Gamma$ belongs to the real interval given by Eq. \eqref{Last Range Gamma}.
Such a peculiar behaviour stems from the interplay between $\Gamma$ and $\theta$ and it appears of relevance since, at least qualitatively, it might give rise to a an experimentally observable effect.
\begin{figure}[tbph]
\centering
{\includegraphics[width=\columnwidth]{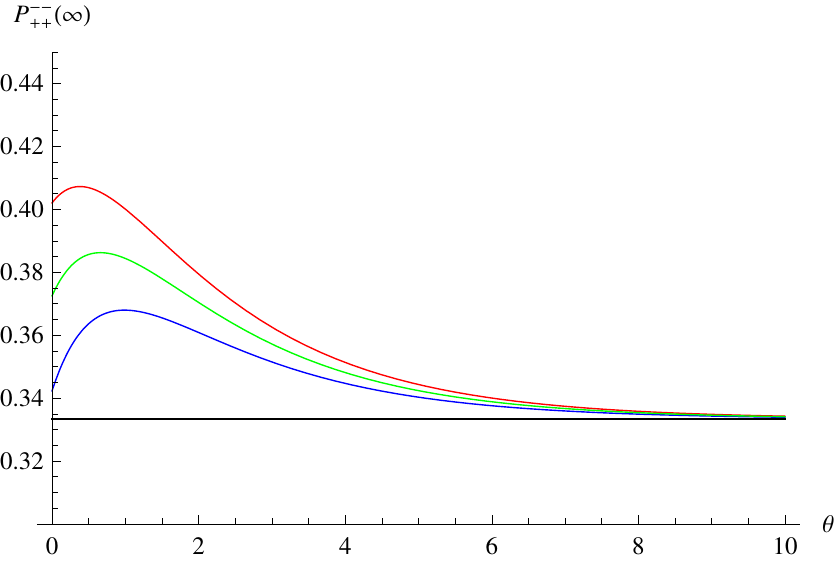} }
\caption{Plot of $P_{++}^{--}(\infty)$ as a function of $\theta$ for different values of $\Gamma$ (from top to bottom): $\Gamma=0.16$, $\Gamma=0.15$, $\Gamma=0.14$; the straight line represents $P_{++}^{--}(\infty)=1/3$.}
\end{figure}

It is interesting to underline that in this physical scenario, the average magnetization of the two-spin 1/2 system at $t=\infty$, $\average{M^z(\infty)} \propto \average{\hat{J}^z(\infty)}=\text{Tr}\{ \rho(\infty) \hat{J}^z \} = P_{++}^{++}-P_{++}^{--}$, with $\hat{J}^z=\hat{j}_1^z+\hat{j}_2^z$, results
\begin{equation}
\average{J^z(\infty)} = e^{-\theta/2}[2e^{-2\pi\Gamma}-1].
\end{equation}
We see that, as expected, the magnetization of the two-spin 1/2 system vanishes in the standard white noise case ($\theta=\infty$), in accordance with the fact that, in this limit condition, in view of Eqs. \eqref{Prob noise ++--} and \eqref{Prob noise ++++}, the populations of the two states $\ket{++}$ and $\ket{--}$ at $t=\infty$ coincide.
The structure of $\average{J^z(\infty)}$ factorized into a function of $\theta$ only and a function of $\Gamma$ only means that the contributions of the noise and of the transverse controllable magnetic field act independently as a consequence of the assumed fast noise scenario \cite{Pokrovsky2}. 
In Fig. 5 we plot $\average{J^z(\infty)}$ as a function of $\Gamma$ for different values of $\theta$.
\begin{figure}[tbph]
\centering
{\includegraphics[width=\columnwidth]{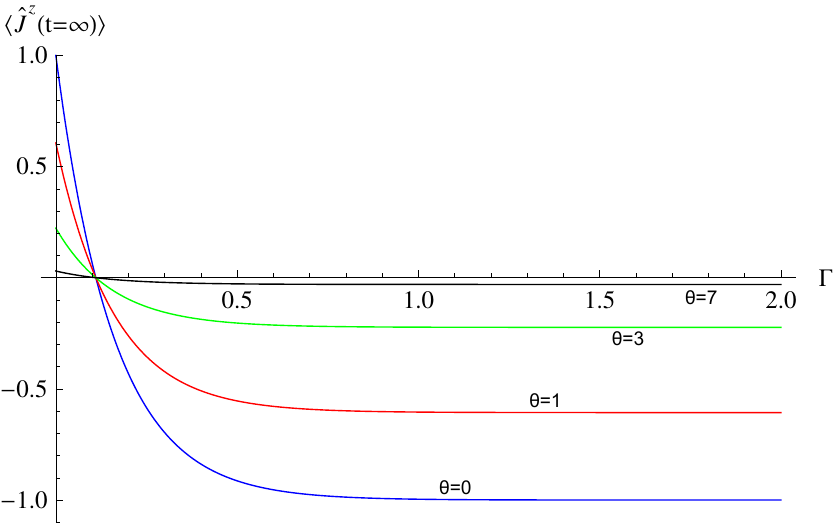} }
\caption{Plots of $\average{J^z(\infty)}$ as a function of $\Gamma$ for different values of $\theta$: $\theta=0$ (blue solid line), $\theta=1$ (red dotted line), $\theta=3$ (green dashed line), $\theta=7$ (black dot-dashed line)}
\end{figure}

For the sake of completeness we report also the transition probabilities when the two-spin 1/2 system is prepared in $\ket{--}$ and $\ket{\Psi^+}$, having respectively
\begin{equation}
P_{--}^{\pm\pm}=P_{++}^{\mp\mp}, \qquad P_{--}^{\pm\mp}=P_{++}^{\pm\mp}, \\
\end{equation}
and
\begin{eqnarray}
&&P_{\Psi^+}^{++}=P_{\Psi^+}^{--}= \\ \nonumber
&&={1 \over 3}\Bigl[1+e^{-3\theta/2}(6e^{-2\pi\Gamma}-6e^{-4\pi\Gamma}-1)\Bigr], \\
&&P_{\Psi^+}^{\Psi^+}=2P_{\Psi^+}^{\pm\mp}= \\ \nonumber
&&={1 \over 3}\Bigl[1-2e^{-3\theta/2}(6e^{-2\pi\Gamma}-6e^{-4\pi\Gamma}-1)\Bigr].
\end{eqnarray}

Our analysis may be repeated for the other scenarios concerning both the noise and the transverse field, exploiting Refs. \cite{Pokrovsky1, Pokrovsky2, Kenmoe1, Kenmoe2}.
We do not proceed further, our goal being to show how the common symmetries of the Hamiltonian models \eqref{Hamiltonian} and \eqref{Ham with noise} are enough to reduce the quantum dynamics of a system of two interacting spins to the quantum dynamics of a single spin both in absence and in presence of noise.
Thus, the main merit of the result reported in this section is that it establishes a qualitative and quantitative link between the dynamical behaviour of a spin $j=1$ subjected to a noisy LZ scenario and the dynamical behaviour of a pair of coupled spin 1/2’s immersed in a noisy LZ scenario too.

\section{Conclusions}\label{CR}

In this paper we have brought to light that the problem 
of a binary system constituted by two, generalized or not, Rabi systems under isotropic Heisenberg 
coupling is reducible into a set of independent problems of single 
(fictitious) spin.
Such property, being based on the structural symmetry imposed to the Hamiltonian model, is immune from effects stemming from degradation of unitary evolution due to the presence of classical random fields and moreover holds whatever the time dependence of the controllable magnetic field is.
We thus claim that this reducibility property, applicable to two quantum spins $\hat{\mathbf{j}}_1$ and $\hat{\mathbf{j}}_2$ of arbitrary magnitudes $j_1$ and $j_2$, represents a new rather general result exploitable in several different physical contexts from condensed matter to quantum information, briefly discussed in the introduction.

Our paper indeed provides ready-for use SU(2)-based expressions of the unitary time evolution operator in terms of the two time-dependent complex-valued functions $a(t)$ and $b(t)$. These two functions determine the joint probability transition of the two spins from an initial state to a final state, in absence of noise.
It is remarkable that under appropriate initial conditions the reduced dynamics of each spin, when noise is ignored, keeps unitarity, meaning that the initial state of the compound system behaves indeed as an IFE state.
The usefulness of our general approach and our results, in absence of noise, have been illustrated for three exemplary cases: two interacting qubits, two interacting qutrits and a qubit interacting with a qutrit.
The time behaviour of the total magnetization as well as of each individual spin (supposed distinguishable) has been exactly forecasted.
It deserves to be emphasized that, in principle and still with no noise, measurements of such time-dependences allow to achieve a feedback both on the coupling mechanism and, if confirmed as Heisenberg exchange interaction, on the coupling constant strength.

The dynamical richness of the adopted Hamiltonian model suggests to investigate on effects stemming from noise.
We have undertaken this task in the last part of our paper.
To this end we have added to the ideal Hamiltonian model a fast fluctuating Gaussian field selected as a classical field, random both in its direction and intensity.
Through our symmetry-based procedure we can write, in the simple case of two spin 1/2's, the joint probability of the LZ transition $P_{++}^{--}$, with the help of the results obtained in Refs. \cite{Pokrovsky2} and \cite{Kenmoe1}.
Our analysis clearly shows that the effects of the fluctuating field on $P_{++}^{--}$ might in general monotonically increase or decrease (Fig. 3) the same quantity we should have in ideal conditions.
Figure 4, on the other hand, transparently illustrates the existence of a maximum in $P_{++}^{--}$ as a function of $\theta$ for special given values of the LZ parameter $\Gamma$.
This result might be qualitatively and experimentally confirmed.


As last remarks, we point out first that it is straightforward to make use of the approach reported in this paper to treat successfully the quantum dynamics of the Hamiltonian model given in Eq. \eqref{Hamiltonian} when the coupling constant $\lambda$ is considered time-dependent too.
Physical scenarios and experimental set-ups leading to time-dependent coupling constants between two subsystems have been recently reported \cite{Anderlini1}.
Secondly, we notice that our approach does not lose its interest even when the experimental set-up in conjunction with the physical system under scrutiny prevent us from invoking distinguishability of two equal and interacting spins.
In this case our approach still holds its validity provided we confine ourselves to any permutationally (symmetric or antisymmetric) invariant subspace $\mathcal{H}^{(j)}$ of the Hilbert space spanned by the eigenstates of the total angular momentum.

\section{Acknowledgements}

Yu.B. was supported by the Government of the Russian Federation (Agreement 05.Y09.21.0018).
H.N. was supported by the Waseda University Grant for Special Research Project (No. 2016B-173).
A.M. and R.G. acknowledge Prof. F. Benatti, Prof. D. Chru\'sci\'nski and Prof. A. S. M. de Castro for interesting and stimulating discussions.
A.M. and R.G. warmly thank also G. Buscarino and M. Todaro for useful comments on possible experimental realizations of the Hamiltonian model and for suggesting Refs. \cite{Bolton,Calvo1,Calvo2}.
R. G. acknowledges too support by research funds in memory of Francesca Palumbo.

\end{document}